\documentclass[usegraphicx,usenatbib]{mn2e}
\usepackage{amsmath}
\usepackage{amssymb}
\usepackage{color}
\usepackage{url}
\usepackage{hyperref}

\usepackage{threeparttable}

\newcommand\lsim{\mathrel{\rlap{\lower4pt\hbox{\hskip1pt$\sim$}}
        \raise1pt\hbox{$<$}}}
\newcommand\gsim{\mathrel{\rlap{\lower4pt\hbox{\hskip1pt$\sim$}}
        \raise1pt\hbox{$>$}}}
\newcommand{\lya}{\ifmmode\mathrm{Ly}\alpha\else{}Ly$\alpha$\fi}
\newcommand{\lyb}{\ifmmode\mathrm{Ly}\beta\else{}Ly$\beta$\fi}
\newcommand{\igm}{\ifmmode\mathrm{IGM}\else{}IGM\fi}
\newcommand{\lae}{\ifmmode\mathrm{LAE}\else{}LAE\fi}
\newcommand{\h}{\ifmmode\mathrm{H}\else{}H\fi}
\newcommand{\hi}{\ifmmode\mathrm{H\,{\scriptscriptstyle I}}\else{}H\,{\scriptsize I}\fi}
\newcommand{\hii}{\ifmmode\mathrm{H\,{\scriptscriptstyle II}}\else{}H\,{\scriptsize II}\fi}
\newcommand{\cmb}{\ifmmode\mathrm{CMB}\else{}CMB\fi}
\newcommand{\qso}{\ifmmode\mathrm{QSO}\else{}QSO\fi}
\newcommand{\eor}{\ifmmode\mathrm{EoR}\else{}EoR\fi}
\newcommand{\heii}{\ifmmode\mathrm{He\,{\scriptscriptstyle II}}\else{}He\,{\scriptsize II}\fi}
\newcommand{\heiii}{\ifmmode\mathrm{He\,{\scriptscriptstyle III}}\else{}He\,{\scriptsize III}\fi}
\newcommand{\ciii}{\ifmmode\mathrm{C\,{\scriptscriptstyle III]}}\else{}C\,{\scriptsize III]}\fi}
\newcommand{\oiii}{\ifmmode\mathrm{O\,{\scriptscriptstyle III}}\else{}O\,{\scriptsize III}\fi}
\newcommand{\aliii}{\ifmmode\mathrm{Al\,{\scriptscriptstyle III}}\else{}Al\,{\scriptsize III}\fi}
\newcommand{\mgii}{\ifmmode\mathrm{Mg\,{\scriptscriptstyle II}}\else{}Mg\,{\scriptsize II}\fi}
\newcommand{\fe}{\ifmmode\mathrm{Fe}\else{}Fe\fi}
\newcommand{\nv}{\ifmmode\mathrm{N\,{\scriptscriptstyle V}}\else{}N\,{\scriptsize V}\fi}
\newcommand{\niv}{\ifmmode\mathrm{N\,{\scriptscriptstyle IV]}}\else{}N\,{\scriptsize IV]}\fi}
\newcommand{\cii}{\ifmmode\mathrm{C\,{\scriptscriptstyle II}}\else{}C\,{\scriptsize II}\fi}
\newcommand{\civ}{\ifmmode\mathrm{C\,{\scriptscriptstyle IV}}\else{}C\,{\scriptsize IV}\fi}
\newcommand{\siv}{\ifmmode\mathrm{Si\,{\scriptscriptstyle IV}}\else{}Si\,{\scriptsize IV}\fi}
\newcommand{\siii}{\ifmmode\mathrm{Si\,{\scriptscriptstyle II}}\else{}Si\,{\scriptsize II}\fi}
\newcommand{\siiii}{\ifmmode\mathrm{Si\,{\scriptscriptstyle III]}}\else{}Si\,{\scriptsize III]}\fi}
\newcommand{\ovi}{\ifmmode\mathrm{O\,{\scriptscriptstyle VI}}\else{}O\,{\scriptsize VI}\fi}
\newcommand{\sioiv}{\ifmmode\mathrm{Si\,{\scriptscriptstyle IV}\,\plus O\,{\scriptscriptstyle IV]}}\else{}Si\,{\scriptsize IV}\,+O\,{\scriptsize IV]}\fi}

\newcommand{\cmmc}{\textsc{\small 21CMMC}}
\newcommand{\cmfst}{\textsc{\small 21cmFAST}}

\pdfoutput=1
\title[SKA Forecasts]{Reionisation \& Cosmic Dawn Astrophysics from the Square Kilometre Array: Impact of Observing Strategies}

\author[B. Greig et al.] {Bradley~Greig$^{1,2}$\thanks{E-mail:~greigb@unimelb.edu.au}, Andrei~Mesinger$^{3}$, \& L\'eon V. E. Koopmans$^{4}$ \\
$^1$ARC Centre of Excellence for All-Sky Astrophysics in 3 Dimensions (ASTRO 3D), University of Melbourne, VIC 3010, Australia \\
$^2$School of Physics, University of Melbourne, Parkville, VIC 3010, Australia \\
$^3$Scuola Normale Superiore, Piazza dei Cavalieri 7, I-56126 Pisa, Italy \\
$^4$Kapteyn Astronomical Institute, University of Groningen, PO Box 800, 9700 AV Groningen, The Netherlands \\
}

\begin{document}
\maketitle \begin{abstract}
\noindent
Interferometry of the cosmic 21-cm signal is set to revolutionise our understanding of the Epoch of Reionisation (EoR) and the Cosmic Dawn (CD).  The culmination of ongoing efforts will be the upcoming Square Kilometre Array (SKA), which will provide tomography of the 21-cm signal from the first billion years of our Universe.  Using a galaxy formation model informed by high-$z$ luminosity functions, here we forecast the accuracy with which the first phase of SKA-low (SKA1-low) can constrain the properties of the unseen galaxies driving the astrophysics of the EoR and CD. We consider three observing strategies: (i) deep (1000h on a single field); (ii) medium-deep (100hr on 10 independent fields); and (iii) shallow (10hr on 100 independent fields). Using the 21-cm power spectrum as a summary statistic, and conservatively {\sl only} using the 21-cm signal above the foreground wedge, we predict that all three observing strategies should recover astrophysical parameters to a fractional precision of $\sim 0.1$ -- 10 per cent.  The reionisation history is recovered to an uncertainty of $\Delta z \lsim 0.1$ (1$\sigma$) for the bulk of its duration.  The medium-deep strategy, balancing thermal noise against cosmic variance, results in the tightest constraints, slightly outperforming the deep strategy. The shallow observational strategy performs the worst, with up to a $\sim 10$ -- 60 per cent increase in the recovered uncertainty. We note, however, that non-Gaussian summary statistics, tomography, as well as unbiased foreground removal would likely favour the deep strategy.
\end{abstract} 

\section{Introduction}

Observing the growth of astrophysical objects (e.g. stars and galaxies) in the first billion years of cosmic history remains elusive. The ubiquity of neutral hydrogen following recombination enshrouds the early Universe in a pervasive, fog rendering it opaque to ultra-violet (UV) light. 
Over time, these primordial galaxies become more abundant and cluster together around high-density peaks until their cumulative output of ionising radiation ionises local (\hii{}) patches of the intergalactic medium (IGM). Percolation of these \hii{} regions through continual star-formation and galaxy growth eventually succeeds in ionising the IGM, referred to as the Epoch of Reionisation (EoR). Unfortunately, the dominant population of sources responsible for reionisation will likely be too faint even for the forthcoming space-based telescopes such as the \textit{James Webb Space Telescope} (JWST) \citep{Gardner:2006p1,Bouwens:2015p7303,Mitra:2015}.

It is not all doom and gloom though. Prior to the completion of reionisation, the sheer abundance of neutral hydrogen will allow us to detect the IGM using the 21-cm spin-flip transition, causing emission or absorption against the Cosmic Microwave Background \citep[see e.g.][]{Gnedin:1997p4494,Madau:1997p4479,Shaver:1999p4549,Tozzi:2000p4510,
  Gnedin:2004p4481,Furlanetto:2006p209,Morales:2010p1274,Pritchard:2012p2958}.
This spatial and frequency (hence also redshift and cosmic time) dependent signal reveals a full three dimensional movie of the IGM during the early Universe.  As it is sensitive to the thermal and ionisation state of the cosmic gas, the 21-cm signal will allow us to infer the typical UV and X-ray properties of the (unseen) galaxy population driving astrophysical processes during the EoR and CD.

However, observing the cosmic 21-cm signal is challenging. It is extremely faint, buried roughly five orders of magnitude below bright astrophysical foregrounds. Nevertheless over the previous decade numerous experiments have sought to statistically detect the signal. These can be broken down into two general categories: (i) large-scale interferometric experiments seeking a measurement of the spatial fluctuations, such as the Murchison Wide Field Array (MWA; \citealt{Tingay:2013p2997}), the Low-Frequency Array (LOFAR; \citealt{vanHaarlem:2013p200,Yatawatta:2013p2980}) and the Precision Array for Probing the Epoch of Reionisation (PAPER; \citealt{Parsons:2010p3000}) and (ii) all-sky averaged global signal experiments, such as the Experiment to Detect the Global EoR Signature (EDGES; \citealt{Bowman:2010p6724}), the Sonda Cosmol\'{o}gica de las Islas para la Detecci\'{o}n de Hidr\'{o}geno Neutro (SCI-HI; \citealt{Voytek:2014p6741}), the Shaped Antenna measurement of the background RAdio Spectrum (SARAS; \citealt{Patra:2015p6814}), Broadband Instrument for Global HydrOgen ReioNisation Signal (BIGHORNS; \citealt{Sokolowski:2015p6827}), the Large Aperture Experiment to detect the Dark Ages (LEDA; \citealt{Greenhill:2012p6829,Bernardi:2016p6834}), Probing Radio Intensity at high-Z from Marion (PRI$^{\rm Z}$M; \citealt{Philip:2019}) and the Netherlands-China Low-Frequency Explorer (NCLE\footnote{https://www.isispace.nl/projects/ncle-the-netherlands-china-low-frequency-explorer/}). 

Aside from an absorption feature in the global signal near $z\approx17$ reported by EDGES \citep{Bowman:2018}, whose interpretation continues to be controversial \citep[see e.g.][]{Hills:2018,Draine:2018,Bowman:2018b,Bradley:2019},
existing experiments have thus far only been able to achieve upper-limits on the 21-cm signal (\citealt{Paciga:2013,Dillon:2015,Jacobs:2015,Beardsley:2016,Patil:2017}; Barry et al., in prep).

Next-generation interferometric experiments, such as the Square Kilometre Array (SKA; \citealt{Mellema:2013p2975}) and the Hydrogen Epoch of Reionization Array (HERA; \citealt{DeBoer:2017p6740}), on the other hand, should be able to achieve higher signal-to-noise measurements of the spatial fluctuations across a broader frequency (redshift) range. Moreover, the SKA will provide the first three-dimensional tomographic image-cubes of the EoR and CD.

In this work, we forecast astrophysical constraints achievable with the SKA1-low\footnote{See \citet{DeBoer:2017p6740} or \citet{Park:2019p9160} for parameter forecasts for HERA.}.   In doing so, we explore several observing strategies, quantifying which one results in the best EoR/CD parameter recovery.\footnote{In this work, we use the 21-cm power spectrum (PS) as a summary statistic when we compute the likelihood of a given set of parameters; however, we note that the sensitivity of the SKA should enable other, non-Gaussian probes of the cosmic 21-cm signal to be detectable \citep[e.g.][]{Watkinson:2014, Yoshiura:2015, Kubota:2016, Shimabukuro:2016, Kakiichi:2017, Shimabukuro:2017, Giri:2018a, Giri:2018b, Majumdar:2018, Gorce:2019, Watkinson:2019} which should further improve our understanding of the astrophysical processes.}
As with any ``beam-steering'' instrument, different observing strategies vary the trade-off between deep/narrow vs shallow/wide observations.  These change the relative balance between the two sources of 21-cm signal measurement errors:
(i) cosmic (sample) variance on large spatial scales and (ii) intrinsic detector (thermal) noise on small spatial scales. Sensitivity on large spatial scales can be improved by increasing the survey volume, while the sensitivity on small scales can be improved by increasing the integration time on a single patch of sky.

What is the optimal trade-off between the two for a fixed amount of observing time?  The conventional approach of judging observing (and foreground removal) strategies by their ability to recover the inputed cosmic 21-cm PS through an integrated signal to noise, assumes that astrophysical insight is encoded in 21-cm fluctuations equally on all scales.
This however, is not the case.  Indeed, moderately-large scales ($k\sim0.1$ Mpc$^{-1}$) seem more sensitive to the properties of the underlying galaxies, compared to small scales \citep[e.g.][]{McQuinn:2007p1665,Greig:2015p3675}.

The remainder of this paper is set up as follows. In Section~\ref{sec:Method} we summarise the astrophysical model used in this analysis as well as the treatment of the instrumental noise and observing strategies. In Section~\ref{sec:results}, we discuss our main finding and in Section~\ref{sec:Conclusion}, we provide our conclusions. Unless stated otherwise, we quote all quantities in co-moving units and adopt the cosmological parameters:  ($\Omega_\Lambda$, $\Omega_{\rm M}$, $\Omega_b$, $n$, $\sigma_8$, $H_0$) = (0.69, 0.31, 0.048, 0.97, 0.81, 68 km s$^{-1}$ Mpc$^{-1}$), consistent with recent results from the Planck mission \citep{PlanckCollaboration:2016p7780}.

\section{Methodology} \label{sec:Method}

\subsection{Simulating the 21-cm signal}

We simulate the cosmic 21-cm signal using the semi-numerical simulation code \cmfst{}\footnote{https://github.com/andreimesinger/21cmFAST}\citep{Mesinger:2007p122,Mesinger:2011p1123}. In particular, we use the most up-to-date astrophysical parameterisation \citep{Park:2019p9160}, which explicitly connects the star-formation rates and ionising escape fraction to the masses of the host dark matter haloes. This step enables \cmfst{}, through some simple conversions, to be able to produce UV luminosity functions (LFs) which can be compared to observed high-$z$ galaxy LFs. Below we briefly summarise \cmfst{} and the astrophysical parameterisation, and refer the reader to these aforementioned works for more details.

\subsubsection{Galaxy UV properties}
We assume that the typical stellar mass of a galaxy, $M_{\ast}$, can be related to its host halo mass, $M_{\rm h}$ \citep[e.g.][]{Kuhlen:2012p1506,Dayal:2014b,Behroozi:2015p1,Mitra:2015,Mutch:2016,Sun:2016p8225,Yue:2016}:
\begin{eqnarray} \label{}
M_{\ast}(M_{\rm h}) = f_{\ast}\left(\frac{\Omega_{\rm b}}{\Omega_{\rm m}}\right)M_{\rm h},
\end{eqnarray}
where $f_{\ast}$ is the fraction of galactic gas in stars which is expressed as a power-law in halo mass,
\begin{eqnarray} \label{}
f_{\ast} = f_{\ast, 10}\left(\frac{M_{\rm h}}{10^{10}\,M_{\odot}}\right)^{\alpha_{\ast}},
\end{eqnarray}
with $f_{\ast, 10}$ being the fraction of galactic gas in stars normalised to a dark matter halo of mass $10^{10}$~$M_{\odot}$ and $\alpha_{\ast}$ is the power-law index.

Next, the star-formation rate (SFR) is estimated by dividing the stellar mass by a characteristic time-scale,
\begin{eqnarray} \label{}
\dot{M}_{\ast}(M_{\rm h},z) = \frac{M_{\ast}}{t_{\ast}H^{-1}(z)},
\end{eqnarray}
where $H^{-1}(z)$ is the Hubble time and $t_{\ast}$ is a free parameter allowed to vary between zero and unity.

The UV ionising escape fraction, $f_{\rm esc}$, is similarly allowed to vary with halo mass,
\begin{eqnarray} \label{}
f_{\rm esc} = f_{\rm esc, 10}\left(\frac{M_{\rm h}}{10^{10}\,M_{\odot}}\right)^{\alpha_{\rm esc}},
\end{eqnarray}
with $f_{\rm esc, 10}$ being normalised to a halo of mass $10^{10}$~$M_{\odot}$.

Finally, we characterise the inability of small mass halos to host active, star-forming galaxies (because of inefficient cooling and/or feedback), through a duty-cycle:
\begin{eqnarray} \label{}
f_{\rm duty} = {\rm exp}\left(-\frac{M_{\rm turn}}{M_{\rm h}}\right).
\end{eqnarray}
In other words, a fraction $(1 - f_{\rm duty})$ of dark matter halos of a mass $M_{\rm h}$ are unable to host star-forming galaxies, with $M_{\rm turn}$ corresponding to the characteristic scale for this suppression \citep[e.g.][]{Shapiro:1994,Giroux:1994,Hui:1997,Barkana:2001p1634,Springel:2003p2176,Mesinger:2008,Okamoto:2008p2183,Sobacchi:2013p2189,Sobacchi:2013p2190}.

\subsubsection{Galaxy X-ray properties}

X-rays from stellar remnants in the first galaxies likely dominate the heating of the IGM, prior to reionisation.
To include the impact of X-ray heating, \cmfst{} computes a cell-by-cell angle-averaged specific X-ray intensity, $J(\boldsymbol{x}, E, z)$, (in erg s$^{-1}$ keV$^{-1}$ cm$^{-2}$ sr$^{-1}$), by integrating the co-moving X-ray specific emissivity, $\epsilon_{\rm X}(\boldsymbol{x}, E_e, z')$ back along the light-cone:
\begin{equation} \label{eq:Jave}
J(\boldsymbol{x}, E, z) = \frac{(1+z)^3}{4\pi} \int_{z}^{\infty} dz' \frac{c dt}{dz'} \epsilon_{\rm X}  e^{-\tau},
\end{equation}
where $e^{-\tau}$ accounts for attenuation by the IGM. The co-moving specific emissivity, evaluated in the emitted frame, $E_{\rm e} = E(1 + z')/(1 + z)$, is, 
\begin{equation} \label{eq:emissivity}
\epsilon_{\rm X}(\boldsymbol{x}, E_{\rm e}, z') = \frac{L_{\rm X}}{\rm SFR} \left[ (1+\bar{\delta}_{\rm nl}) \int^{\infty}_{0}{\rm d}M_{\rm h} \frac{{\rm d}n}{{\rm d}M_{\rm h}}f_{\rm duty} \dot{M}_{\ast}\right],
\end{equation}
where $\bar{\delta}_{\rm nl}$ is the mean, non-linear density in a shell around $(\boldsymbol{x}, z)$ and the quantity in square brackets is the SFR density along the light-cone. 

The normalisation, $L_{\rm X}/{\rm SFR}$ (erg s$^{-1}$ keV$^{-1}$ $M^{-1}_{\odot}$ yr), is the specific X-ray luminosity per unit star formation escaping the host galaxies. It is assumed that the specific intensify follows a power-law with respect to photon energy, $L_{\rm X} \propto E^{- \alpha_X}$, with photons below a threshold energy, $E_0$, being absorbed inside the host galaxy\footnote{For this work, we assume a fixed power-law slope of $\alpha_X = 1$ consistent with observations of high-mass X-ray binaries \citep{Mineo:2012p6282,Fragos:2013p6529,Pacucci:2014p4323}.
}. This specific luminosity is then normalised to the integrated soft-band ($<2$~keV) luminosity per SFR (in erg s$^{-1}$ $M^{-1}_{\odot}$ yr), which we take to be a free parameter:
\begin{equation} \label{eq:normL}
  L_{{\rm X}<2\,{\rm keV}}/{\rm SFR} = \int^{2\,{\rm keV}}_{E_{0}} dE_e ~ L_{\rm X}/{\rm SFR} ~.
\end{equation}
This limit of $2\,{\rm keV}$ equates to roughly the Hubble length at high redshifts, implying that harder photons do not heat the IGM \citep[e.g.][]{McQuinn:2012p3773}.

\subsubsection{Computing the 21-cm signal}

The 21-cm signal is commonly expressed in terms of a brightness temperature contrast with respect to the Cosmic Microwave Background (CMB) temperature, $T_{\rm CMB}$ \citep[e.g.][]{Furlanetto:2006p209}:
\begin{eqnarray} \label{eq:21cmTb}
\delta T_{\rm b}(\nu) &=& \frac{T_{\rm S} - T_{\rm CMB}(z)}{1+z}\left(1 - {\rm e}^{-\tau_{\nu_{0}}}\right)~{\rm mK},
\end{eqnarray}
where $\tau_{\nu_{0}}$ is the optical depth of the 21-cm line, which is:
\begin{eqnarray}
\tau_{\nu_{0}} &\propto& (1+\delta_{\rm nl})(1+z)^{3/2}\frac{x_{\hi{}}}{T_{\rm S}}\left(\frac{H}{{\rm d}v_{\rm r}/{\rm d}r+H}\right)
\end{eqnarray}
Here, $x_{\hi{}}$ is the neutral hydrogen fraction, $\delta_{\rm nl} \equiv \rho/\bar{\rho} - 1$ is the gas over-density, $H(z)$ is the Hubble parameter,  ${\rm d}v_{\rm r}/{\rm d}r$ is the gradient of the line-of-sight component of the velocity and $T_{\rm S}$ is the gas spin temperature. All quantities are evaluated at redshift $z = \nu_{0}/\nu - 1$, where $\nu_{0}$ is the 21-cm frequency and we drop the spatial dependence for brevity.

\cmfst{} generates evolved density and velocity fields using second-order Lagrange perturbation theory \citep[e.g][]{Scoccimarro:1998p7939} from high resolution Gaussian initial conditions. Reionisation is computed from the evolved density field by comparing the cumulative number of ionising photons to the number of neutral hydrogen atoms plus cumulative recombinations in spheres of decreasing radii. At each cell, ionisation occurs when,
\begin{eqnarray} \label{eq:ioncrit}
n_{\rm ion}(\boldsymbol{x}, z | R, \delta_{R}) \geq (1 + \bar{n}_{\rm rec})(1-\bar{x}_{e}),
\end{eqnarray}
where $\bar{n}_{\rm rec}$ is the cumulative number of recombinations \citep[e.g.][]{Sobacchi:2014p1157} and $n_{\rm ion}$ is the cumulative number of IGM ionising photons per baryon inside a spherical region of size, $R$ and corresponding overdensity, $\delta_{R}$,
\begin{eqnarray} \label{eq:ioncrit2}
n_{\rm ion} = \bar{\rho}^{-1}_b\int^{\infty}_{0}{\rm d}M_{\rm h} \frac{{\rm d}n(M_{h}, z | R, \delta_{R})}{{\rm d}M_{\rm h}}f_{\rm duty} \dot{M}_{\ast}f_{\rm esc}N_{\gamma/b},
\end{eqnarray}
where $\rho_b$ is the mean baryon density and $N_{\gamma/b}$ is the number of ionising photons per stellar baryon\footnote{We take this number to be 5000, corresponding to a Salpeter initial mass function \citep{Salpeter:1955}; however this is highly degenerate with $f_{\ast}$}.
The final term of Equation~\ref{eq:ioncrit}, $(1-\bar{x}_{e})$, corresponds to the number of ionisations by X-rays, expected to contribute at a level of less than $\sim 10$ per cent \citep[e.g.][]{Ricotti:2004p7145,Mesinger:2013p1835,Madau:2017,Ross:2017,Eide:2018}

The temperature and the level of partial ionisation of the neutral IGM is tracked in each cell, accounting for adiabatic heating/cooling, Compton heating/cooling, heating through partial ionisations, as well as the heating/ionisations from X-rays (discussed in the previous section).  The spin temperature is then computed as a weighted mean between the gas and CMB temperatures, depending on the density and local Lyman-$\alpha$ intensity impinging on each cell \citep{Wouthuysen:1952p4321,Field:1958p1}.

Finally, we combine all the cosmological fields to compute the cosmic 21-cm signal, as outlined in Equation~\ref{eq:21cmTb}. Additionally, we include the impact of redshift space distortions along the line-of-sight as outlined in \citet{Mao:2012p7838,Jensen:2013p1389,Greig:2018p3217}.

\subsection{Astrophysical parameter set} \label{sec:fiducial}

Under the assumption of this astrophysical model, we are left with eight free parameters, which we summarise below. We adopt the same fiducial model and allowed parameter ranges from \citet{Park:2019p9160}. This model is summarised in Table~\ref{tab:Results}, and its parameters are:
\begin{itemize}
\item[(i)]$f_{\ast, 10}$: normalisation for the fraction of galactic gas in stars evaluated at a halo mass of 10$^{10}~M_{\odot}$. We adopt a fiducial model of $f_{\ast, 10} = 0.05$ and vary the log quantity as ${\rm log}_{10}(f_{\ast, 10}) \in [-3,0]$.
\item[(ii)]$\alpha_{\ast}$: power-law index for the star-formation as a function of halo mass. We adopt a fiducial value of $\alpha_{\ast} = 0.5$, allowing it to vary in the range $\alpha_{\ast} \in [-0.5,1]$.
\item[(iii)]$f_{\rm esc, 10}$: normalisation for the ionising UV escape fraction evaluated at a halo mass of 10$^{10}~M_{\odot}$. We adopt $f_{\rm esc, 10} = 0.1$ to be our fiducial value, allowing it to vary in the range $f_{\rm esc, 10} \in [-3, 0]$.
\item[(iv)]$\alpha_{\rm esc}$: power-law index for the ionising UV escape fraction as a function of halo mass. We adopt a fiducial value of $\alpha_{\ast} = -0.5$, allowing it to vary in the range $\alpha_{\ast} \in [-1,0.5]$.
\item[(v)]$t_{\ast}$: the star-formation time scale as a fraction of the Hubble time. Fiducially, we adopt $t_{\ast} = 0.5$ allowing it to vary in the range $t_{\ast} \in (0,1]$.
\item[(vi)]$M_{\rm turn}$: halo mass turn-over below which the abundance of active star-forming galaxies is exponentially suppressed by the adopted duty cycle. We adopt $M_{\rm turn} = 5\times10^8$~$M_{\odot}$ to be our fiducial choice, with it being allowed to vary within the range ${\rm log}_{10}(M_{\rm turn}) \in[8,10]$.
\item[(vii)]$E_{0}$: the minimum energy threshold for X-ray photons capable of escaping their host galaxy. We adopt a fiducial value of $E_{0} = 0.5$~keV allowing it to vary within the range $E_{0} \in [0.2,1.5]$~keV. For reference, this corresponds of a integrated column density of ${\rm log_{10}}(N_{\hi{}}/{\rm cm}^{2}) \in [19.3,23.0]$.
\item[(viii)]$L_{{\rm X}<2\,{\rm keV}}/{\rm SFR}$: the normalisation for the soft-band X-ray luminosity per unit star-formation determined over the $E_{0} - 2$~keV energy band. Fiducially we adopt a value of ${\rm log_{10}}(L_{{\rm X}<2\,{\rm keV}}/{\rm SFR}) = 40.5$, and allow it to vary in the range ${\rm log_{10}}(L_{{\rm X}<2\,{\rm keV}}/{\rm SFR}) \in [38, 42]$.
\end{itemize}

\subsection{Modelling the astrophysical noise}

As we aim to explore the performance of a variety of observing strategies for the SKA1--low, we must be able to model the expected instrumental noise. Since we focus on the 21-cm PS, we use the publicly available \textsc{\small Python} module \textsc{\small 21cmSense}\footnote{https://github.com/jpober/21cmSense}\citep{Pober:2013p41,Pober:2014p35} and briefly summarise the method below.

The thermal noise PS is estimated by gridding the $uv$-visibilities according to \citep[e.g.][]{Morales:2005p1474,McQuinn:2006p109,Pober:2014p35},
\begin{eqnarray} \label{eq:NoisePS}
\Delta^{2}_{\rm N}(k) \approx X^{2}Y\frac{k^{3}}{2\pi^{2}}\frac{\Omega^{\prime}}{2t}T^{2}_{\rm sys},
\end{eqnarray} 
where $X^{2}Y$ converts between observing bandwidth, frequency and co-moving distance, $\Omega^{\prime}$ is a 
beam-dependent factor derived in \citet{Parsons:2014p781}, $t$ is the total time spent by all baselines within a particular $k$-mode and 
$T_{\rm sys}$ is the system temperature, the sum of the receiver temperature, $T_{\rm rec}$, and the sky temperature $T_{\rm sky}$. 
We model $T_{\rm sky}$ using the frequency dependent scaling $T_{\rm sky} = 60\left(\frac{\nu}{300~{\rm MHz}}\right)^{-2.55}~{\rm K}$ \citep{Thompson2007}.

The sample (cosmic) variance contribution to the error on the inferred PS is estimated from a cosmological 21 cm PS (i.e. our fiducial mock observation of the 21-cm PS, $\Delta^{2}_{21}(k)^{2}$) and is combined with the thermal noise using an inverse-weighted summation over all the individual modes \citep{Pober:2013p41}. This results in a total noise power, $\delta\Delta^{2}_{\rm T+S}(k)$, at a given Fourier mode, $k$,
\begin{eqnarray} \label{eq:T+S}
\delta\Delta^{2}_{\rm T+S}(k) = \left(\sum_{i}\frac{1}{(\Delta^{2}_{{\rm N},i}(k) + \Delta^{2}_{21}(k))^{2}}\right)^{-\frac{1}{2}}.
\end{eqnarray}
Inherently, this assumes Gaussian errors for the cosmic-variance term, which for most scales is a relatively good approximation (though see \citealt{Mondal:2015p4936,Shaw:2019p08706} for more detailed discussions).  

Finally, we adopt the conservative ``moderate'' foreground treatment from \citet{Pober:2014p35}. This constitutes foreground avoidance, where we restrict the computation of the 21-cm PS to modes outside of the contaminated foreground ``wedge''.

\subsection{SKA design and observing strategies} \label{sec:SKA}

We estimate the SKA1--low sensitivity curves using the antennae station layout according to the recent SKA System Baseline Design document\footnote{http://astronomers.skatelescope.org/wp-content/uploads/2016/09/SKA-TEL-SKO-0000422\textunderscore 02\textunderscore SKA1\textunderscore LowConfigurationCoordinates-1.pdf}. This consists of 512 35m antennae stations randomly distributed within a 500m core radius. The total system temperature is modelled as $T_{\rm sys} = 1.1T_{\rm sky} + 40~{\rm K}$. SKA1--low is a phase-tracking experiment, for which we assume that we can conservatively perform a single six-hour track per night. 

In this work, we want to explore the performance of various observing strategies for SKA. To do this, we assume a fixed survey footprint, corresponding to a total integration time of 1000hr. In principle, with the multi-beaming capabilities of the SKA one could obtain two fields per observation (i.e. 2 independent 1000hr fields for the total time cost of 1000hr), however, we restrict our analysis to a single pointing for simplicity\footnote{In practise, the nominal planned survey for the SKA (the deep survey) will cover $\sim100$~deg$^{2}$ requiring 2500~hr on sky in dual-beam mode \citep{Koopmans:2015}.}. With 1000hrs of integration time, we consider three possible observing strategies:
\begin{itemize}
\item[(i)] deep (1000hr) : A single, deep 1000hr integration of a $\sim20$~deg$^2$ (at 150 MHz) cold patch of sky. The SKA is primarily an imaging experiment for the EoR, thus to perform a tomographic study of the 21-cm signal the thermal noise must be minimised at the expense of cosmic variance. Thus, this strategy will be most sensitive to small spatial scales (large $k$-modes).
\item[(ii)] medium-deep ($10\times100$hr): A balance between cosmic variance and thermal noise. We observe 10 independent patches of the sky for an intermediate 100hrs.
\item[(iii)] Shallow ($100\times10$hr): A shallow, but wide survey observing 100 independent patches of the sky. Minimises the cosmic variance, reducing the noise on large scales (small $k$).
\end{itemize}

We note that the transformative power of the SKA will be in performing 21-cm tomography (i.e. direct imaging of the 21-cm signal) of the first billion years.  For this measurement, it is important to have a good $uv$-coverage, and a high signal-to-noise.  Thus, regardless of its performance in parameter recovery using the PS, the deep field observation will be optimal for imaging of the 21-cm signal.
This will also allow us to characterise the cosmic signal with non-Gaussian statistics (something we do not investigate here).

\subsection{\cmmc{} setup}

\cmmc{} is a massively parallel Monte-Carlo Markov Chain (MCMC) sampler of 3D semi-numerical reionisation simulations \citep{Greig:2015p3675,Greig:2017p8496,Greig:2018p3217,Park:2019p9160}. It is based off the \textsc{\small Python} module \textsc{\small CosmoHammer} \citep{Akeret:2012p842} which uses the \textsc{\small Emcee} \textsc{\small Python} module \citep{ForemanMackey:2013p823}, an affine invariant ensemble sampler from \citet{Goodman:2010p843}. At each proposal step, \cmmc{} performs an independent 3D realisation of the 21-cm signal using \cmfst{} to obtain a sampled 21-cm PS. A likelihood is then estimated by comparing this sampled PS against a mock (input) PS. We calculate this likelihood over a limited $k$-space range of $k=0.1-1.0$~Mpc$^{-1}$, where the lower limit is set by noise from astrophysical foregrounds while the upper limit is set by shot noise from the resolution of the simulations, respectively.

In addition to instrumental noise, we include two other sources of uncertainty. First, we adopt an uncorrelated, multiplicative modelling uncertainty of 20 per cent applied to the sampled 21-cm PS. This is motivated by approximations adopted in semi-numerical simulations relative to radiative-transfer simulations \citep[e.g.][]{Zahn:2011p1171,Ghara:2018,Hutter:2018}. Second, we include Poisson errors on the sampled PS roughly consistent with sample variance on these scales. These two sources of uncertainty are then combined with the total noise PS from Equation~\ref{eq:T+S} by summing in quadrature.

In order to provide our astrophysical parameter forecasts we must construct a mock observation from which we aim to recover the input parameter values. Using the fiducial parameters outlined in Section~\ref{sec:fiducial} we construct a mock 21-cm light-cone, with a transverse scale of 500~Mpc and 256 voxels per side length. For the MCMC itself, we then sample 3D realisations of the 21-cm light-cone with a transverse scale of 250~Mpc and 128 voxels per side length. To perform the likelihood calculation, we split the 21-cm light-cone into equal 250~Mpc comoving-volume depths within which we calculate the 3D spherically averaged 21-cm PS. This results in twelve 21-cm PS which span the SKA1--low frequency bandwidth, $z\sim6-27$ (50--200~MHz).

In combination with the 21-cm PS from our mock observation, we additionally include priors from high redshift galaxy LFs.
Following \citet{Park:2019p9160} we use the $z\sim6$ LF from \citet{Bouwens:2017}, $z\sim7-8$ from \citet{Bouwens:2015} and $z\sim10$ from \citet{Oesch:2018}. Including these priors enables us to improve the constraining power on the astrophysical parameterisation used in this work because it breaks degeneracies amongst parameters less sensitive to the 21-cm signal (e.g. the star-formation time scale, $t_{\ast}$, see \citealt{Park:2019p9160} for more in-depth discussions).

\section{Observing-Strategy Forecasts} \label{sec:results}

\begin{figure*} 
	\begin{center}
	  \includegraphics[trim = 0.7cm 1.7cm 0cm 0.7cm, scale = 0.6]{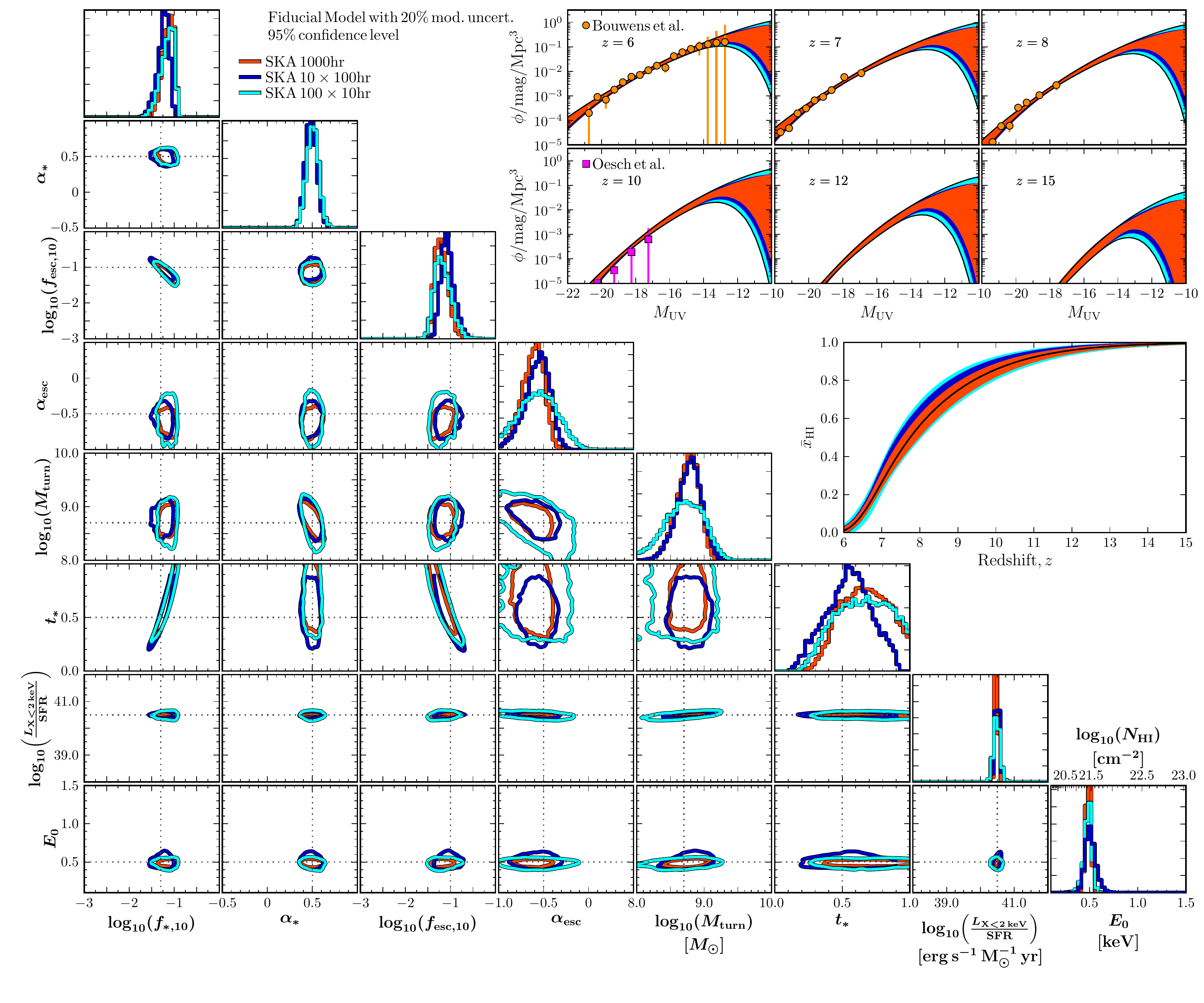}
	\end{center}
\caption[]{Recovered one and two dimensional marginalised contours for the astrophysical parameters for our three different observing strategies with the SKA: (i) 1000hr (deep) -- red (ii) $10\times100$~hr (medium-deep) -- blue and (iii) $100\times10$~hr (shallow) -- cyan. For all, we include the 20 per cent modelling uncertainty. Black dotted lines correspond to the input fiducial model parameters. Top right panels are the recovered 95 percentiles on the UV LFs at several redshifts compared to the input observed LFs (used as observational priors and represented by the orange and pink data points). Middle right corresponds to the global evolution of the IGM neutral fraction, $\bar{x}_{\hi{}}$.}
\label{fig:DiffStrat}
\end{figure*}

In Figure~\ref{fig:DiffStrat}, we present the recovered one and two dimensional marginalised constraints for our input astrophysical model as well as the recovered UV LFs and the global evolution of the IGM neutral fraction, $\bar{x}_{\hi{}}$. Additionally, in Table~\ref{tab:Results} we provide the marginalised 68th percentiles for each astrophysical parameter. These correspond to the main results of this work. For reference we additionally include the astrophysical parameter constraints for a 1000hr observation with HERA from \citet{Park:2019p9160}.

\subsection{Comparing observing strategies}

\begin{table*}
\tiny
\begin{tabular}{@{}lccccccccc}
\hline
  & ${\rm log_{10}}(f_{\ast,10})$ & $\alpha_{\ast}$ & ${\rm log_{10}}(f_{\rm esc,10})$ & $\alpha_{\rm esc}$ & $t_{\ast}$ & ${\rm log_{10}}(M_{\rm turn})$ & ${\rm log_{10}}\left(\frac{L_{{\rm X}<2{\rm keV}}}{\rm SFR}\right)$ & $E_0$   \\
               &  &  &  &  & & $[{\rm M_{\sun}}]$ & $[{\rm erg\,s^{-1}\,M_{\sun}^{-1}\,yr}]$ &  $[{\rm keV}]$ \\
\hline
\vspace{0.8mm}
Mock Obs. & $-1.30$ & $0.50$ & $-1.00$ & $-0.50$ & $0.5$ & $8.7$ & $40.50$ &  $0.50$\\
\hline
\vspace{0.8mm}
HERA 331 (1000hr) & $-1.20\substack{+0.14 \\ -0.14}$ &  $0.47\substack{+0.06 \\ -0.06}$  &  $-1.10\substack{+0.16 \\ -0.18}$  &  $-0.48\substack{+0.14 \\ -0.18}$  &  $0.56\substack{+0.21 \\ -0.16}$ &   $8.76\substack{+0.19 \\ -0.23}$   &   $40.49\substack{+0.05 \\ -0.06}$   &   $0.50\substack{+0.03 \\ -0.03}$\\
\hline
\vspace{0.8mm}
SKA (1000hr) & $-1.12\substack{+0.11 \\ -0.15}$ &  $0.49\substack{+0.06 \\ -0.06}$  &  $-1.21\substack{+0.16 \\ -0.13}$  &  $-0.61\substack{+0.11 \\ -0.13}$   &  $0.67\substack{+0.19 \\ -0.19}$ &   $8.77\substack{+0.15 \\ -0.19}$   &   $40.48\substack{+0.04 \\ -0.04}$   &   $0.49\substack{+0.03 \\ -0.03}$\\
\vspace{0.8mm}
SKA (10x100hr) & $-1.14\substack{+0.11 \\ -0.15}$ &  $0.48\substack{+0.06 \\ -0.06}$  &  $-1.17\substack{+0.16 \\ -0.13}$  &  $-0.56\substack{+0.12 \\ -0.15}$     &  $0.64\substack{+0.17 \\ -0.18}$ &   $8.79\substack{+0.17\\ -0.19}$   &   $40.50\substack{+0.06 \\ -0.06}$   &   $0.51\substack{+0.06 \\ -0.04}$\\
\vspace{0.8mm}
SKA (100x10hr) & $-1.14\substack{+0.12 \\ -0.18}$ &  $0.50\substack{+0.06 \\ -0.07}$  &  $-1.18\substack{+0.20 \\ -0.16}$  &  $-0.56\substack{+0.21 \\ -0.22}$  &  $0.66\substack{+0.21 \\ -0.22}$  &   $8.71\substack{+0.28 \\ -0.30}$   &   $40.49\substack{+0.09 \\ -0.09}$   &   $0.49\substack{+0.04 \\ -0.04}$\\
\hline
\vspace{0.8mm}
No Modelling Uncertainty  \\
SKA (10x100hr) & $-1.23\substack{+0.09 \\ -0.11}$ &  $0.47\substack{+0.05 \\ -0.06}$  &  $-1.11\substack{+0.12 \\ -0.11}$  &  $-0.56\substack{+0.09 \\ -0.12}$   &  $0.56\substack{+0.15 \\ -0.15}$ &   $8.72\substack{+0.14 \\ -0.15}$    &   $40.48\substack{+0.05 \\ -0.05}$   &   $0.49\substack{+0.05 \\ -0.04}$\\
\hline
\end{tabular}
\caption{Summary of the recovered precision (68 percentiles) for all the astrophysical parameters considered in this work. These include recovery of the 21-cm PS from a mock observation (parameters in top row) with observed UV LFs as an input prior. For comparison, we include the expected constraints for HERA as generated in \citet{Park:2019p9160}.}
\label{tab:Results}
\end{table*} 

Surprisingly, for the majority of the astrophysical parameters, all three observing strategies perform equally well. Note that for the X-ray parameters, there is no distinguishable difference between the strategies. However, for $\alpha_{\rm esc}$, $M_{\rm turn}$ and $t_{\ast}$ the largest differences occur. Thus, the observing strategies have the largest impact on the galaxy UV properties. Firstly, it is immediately obvious that the shallow ($100\times10$~hr) survey incurs the largest errors. This is equally reflected in the broader recovered UV LFs and reionisation history. Clearly, by focussing on the largest scales (smallest $k$-modes), constraining information is lost from the intermediate to smaller scales (larger $k$-modes) where thermal noise dominates. Note though that we restrict our likelihood fitting to $k=0.1 - 1.0$~Mpc$^{-1}$. If this lower bound could be reduced (requiring observing into the foreground wedge) the relative performance of the shallow survey would be improved
as it is most sensitive to modes within the foreground `wedge'.

On the other hand, the deep (1000hr) and medium-deep ($10\times100$~hr) surveys result in comparable constraints.
To highlight these similarities, we cast the 68th percentiles as approximate $1\sigma$ uncertainties. In doing so, we find 
[${\rm log}_{10}(f_{\ast, 10})$, $\alpha_{\ast}$, ${\rm log}_{10}(f_{\rm esc, 10})$, $\alpha_{\rm esc}$, $t_{\ast}$, ${\rm log}_{10}(M_{\rm turn})$, $E_{0}$, ${\rm log}_{10}(L_{{\rm X}<2\,{\rm keV}}/{\rm SFR})$] = (11.6, 12.2, 12.0, 19.7, 28.4, 1.9, 0.1, 6.1) per cent for the deep scenario and (11.4, 12.3, 12.4, 24.0, 27.3, 2.0, 0.1, 9.8) per cent for the medium-deep scenario.

For the star-formation timescale, $t_{\ast}$, the medium-deep strategy recovers notably tighter constraints as highlighted by the one-dimensional marginalised histogram for $t_{\ast}$. However, the deep survey strategy recovers marginally tighter UV LFs and reionisation history. 
These differences arise mostly from the degeneracies between $t_{\ast}$--$f_{\rm esc,10}$ and $t_{\ast}$--$f_{\rm \ast,10}$. Notably, for the deep survey all three quantities are slightly offset from their expected fiducial value unlike that for the medium-deep strategy. This slight offset in these parameters from the deep survey in combination with the tighter $t_{\ast}$ constraints for the medium-deep scenario indicates that the medium-deep strategy is the preferred observing strategy. The source of this slight offset likely arises from two correlated sources: (i) how the sensitivity for each strategy is distributed over $k$-space for the 21-cm PS and (ii) that the fiducial galaxy UV parameters were not \textit{a priori} selected to be an exact match to the input observational priors (UV LFs). For the former, the deep strategy prefers the smallest scales (large $k$-modes), whereas the medium-deep scenario pushes further into the large-scale modes. We anticipate the largest scales to be the most sensitive to the astrophysical information, thus a more uniform distribution of noise from the medium-deep strategy over $k$-space will improve the astrophysical parameter recovery. For the latter, the best recovered UV parameters from the UV LFs alone differ from those recovered from the 21-cm PS. Coupling this with differences in the sensitivity per $k$-mode will cause slight offsets when the 21-cm PS recovery is less sensitive as shown in \citet{Park:2019p9160}.

It is important to remember here that we only use the 21-cm PS to compute the likelihood. Additionally including non-Gaussian statistics \citep[e.g.][]{Watkinson:2014, Yoshiura:2015, Kubota:2016, Shimabukuro:2016, Kakiichi:2017, Shimabukuro:2017, Shimabukuro:2017b, LaPlante:2018, Majumdar:2018, Giri:2018a, Giri:2018b, Gillet:2019, Gorce:2019, Hassan:2019, Watkinson:2019} would improve the relative performance of the deep survey, as it has the largest signal-to-noise for higher order statistics.

\subsection{Comparison to HERA}

Finally, in Table~\ref{tab:Results}, we compare the astrophysical forecasts from SKA to those for a 1000hr observation from HERA explored in \citet{Park:2019p9160}. Note, in both instances we only use the PS space above the wedge. Again, we caution that this is not a direct like-for-like comparison as HERA is a drift scan observation compared to the tracked scanning to be performed by the SKA. Thus, HERA will have better sensitivity on larger scales owing to reduced sample variance (i.e. more independent observing fields) at the expense of small-scale sensitivity owing to increased thermal noise. Nevertheless, we find that the medium-deep observing strategy marginally outperforms HERA as evidenced by the slightly reduced fractional errors on the recovered astrophysical parameters. We note, however, that SKA aims to remove the foregrounds \citep{Koopmans:2015} and utilise the full PS space inside the wedge as well, potentially significantly increasing its power to recover astrophysical parameters \citep{DeBoer:2017p6740}. Approximating these percentiles as $1\sigma$ fractional errors we find
[${\rm log}_{10}(f_{\ast, 10})$, $\alpha_{\ast}$, ${\rm log}_{10}(f_{\rm esc, 10})$, $\alpha_{\rm esc}$, $t_{\ast}$, ${\rm log}_{10}(M_{\rm turn})$, $E_{0}$, ${\rm log}_{10}(L_{{\rm X}<2\,{\rm keV}}/{\rm SFR})$] = (11.4, 12.3, 12.4, 24.0, 27.3, 2.0, 0.1, 9.8) per cent for the SKA and (11.7, 12.8, 15.5, 33.3, 33.0, 2.4, 0.1, 6.0) for HERA. Note again though that offsets arise in the median recovered astrophysical parameters relative to the fiducial parameters. However, these again can be attributed to the combined effect of the instrumental sensitivity on different $k$-scales and the chosen input UV LFs not preferring the same fiducial galaxy UV parameters as the mock observation. 
 
\subsection{Impact of modelling uncertainty}

\begin{figure*} 
	\begin{center}
	  \includegraphics[trim = 0.7cm 1.7cm 0cm 0.7cm, scale = 0.6]{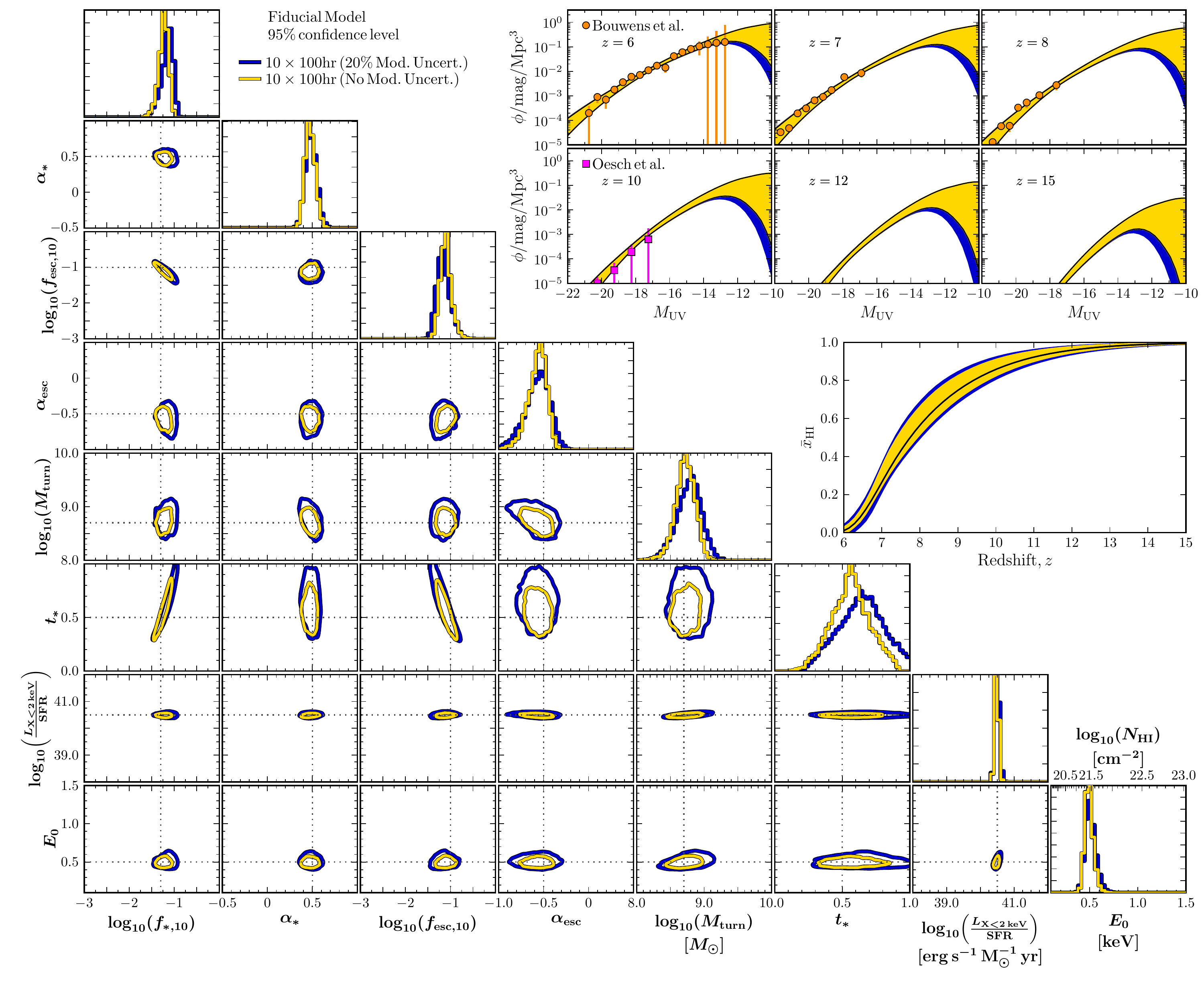}
	\end{center}
\caption[]{Same as Figure~\ref{fig:DiffStrat} except now we compare the impact of the modelling uncertainty on the $10\times100$~hr (medium-deep) observing strategy: (i) 20 per cent modelling uncertainty -- blue and (ii) no modelling uncertainty -- yellow.}
\label{fig:ModUncert}
\end{figure*}

Throughout this work we have included an additional 20 per cent modelling uncertainty to our estimation of the likelihood. However, it is useful to explore the idealised case that modelling errors can be efficiently characterised and accounted for.
Thus in Figure~\ref{fig:ModUncert} and summarised at the bottom of Table~\ref{tab:Results}, we compare our best-performing observing strategy, the medium-deep survey, with and without this modelling uncertainty.

Approximating the marginalised PDFs to obtain simplified $1\sigma$ fractional errors, we find [${\rm log}_{10}(f_{\ast, 10})$, $\alpha_{\ast}$, ${\rm log}_{10}(f_{\rm esc, 10})$, $\alpha_{\rm esc}$, $t_{\ast}$, ${\rm log}_{10}(M_{\rm turn})$, $E_{0}$, ${\rm log}_{10}(L_{{\rm X}<2\,{\rm keV}}/{\rm SFR})$] = (11.4, 12.3, 12.4, 24.0, 27.3, 2.0, 0.1, 9.8) per cent for the medium-deep scenario with the modelling uncertainty compared to (8.1, 11.7, 10.4, 18.8, 26.8, 1.7, 0.1, 9.2) without the modelling uncertainty. Thus, including a 20 per cent modelling uncertainty increases the fractional uncertainties by (40.7, 5.1, 19.2, 27.7, 1.9, 18.0, 20.0, 6.5) per cent. The largest improvement in the recovery is for the star-formation time-scale parameter.  However, in general, assuming no modelling error does not improve the recovery dramatically, suggesting that it is not the largest source of uncertainty, for this mock observation.

\section{Conclusion} \label{sec:Conclusion}

The ultimate goal for current and future reionisation experiments is to recover a full three-dimensional view of the early Universe through detection of the 21-cm signal of neutral hydrogen. In doing so, we will be able to obtain insights into the formation and nature of the first stars and galaxies along with their growth over the first billion years.

The most ambitious upcoming 21-cm telescope is the SKA. Here we provide EoR/CD astrophysical parameter forecasts achievable with SKA1-low under some very conservative assumptions: (i) that only the EoR window above the foreground wedge is used and (ii) 20 per cent modelling uncertainties are included. We use a physically-motivated galaxy formation model which allows us to make use of observed LFs of high-$z$ galaxies, in addition to mock SKA 21-cm PS measurements.

We consider three different SKA observing strategies, quantifying the trade-off
between minimising the errors associated with cosmic (sample) variance and instrumental (thermal) noise. For a fixed total integration time, we considered: (i) a deep 1000hr observation of a single patch of sky (ii) a medium-deep 100hr observation of 10 independent fields and (iii) a shallow 10hr observation of 100 independent fields. We note that the SKA aims to observe about five times this volume \citep{Koopmans:2015}.

Under the above assumptions, we find that the deep and medium-deep observing strategies perform almost equally well, both yielding tighter parameter constraints compared with the shallow strategy.
Approximated as $1\sigma$ uncertainties the medium-deep survey recovers the following constraints: [${\rm log}_{10}(f_{\ast, 10})$, $\alpha_{\ast}$, ${\rm log}_{10}(f_{\rm esc, 10})$, $\alpha_{\rm esc}$, $t_{\ast}$, ${\rm log}_{10}(M_{\rm turn})$, $E_{0}$, ${\rm log}_{10}(L_{{\rm X}<2\,{\rm keV}}/{\rm SFR})$] = (11.4, 12.3, 12.4, 24.0, 27.3, 2.0, 0.1, 9.8) per cent.

Additionally, we explore the impact of our chosen 20 per cent modelling uncertainty on our recovered astrophysical parameters. We find that an optimistic scenario in which the modelling error can be completely corrected for, only modestly improves parameter constraints (at most tens of per cent). Thus, a modelling error at the level of a few tens of per cent does not strongly degrade the accuracy of parameter recovery, for our galaxy formation model. With SKA1-low we therefore will be able to recover the astrophysics of reionisation and the CD at the level of $\sim$ 10 per cent, or better.

\section*{Acknowledgements}

Parts of this research were supported by the Australian Research Council Centre of Excellence for All Sky Astrophysics in 3 Dimensions (ASTRO 3D), through project number CE170100013. AM acknowledges funding support from the European Research Council (ERC) under the European Union's Horizon 2020 research and innovation programme (grant agreement No 638809 -- AIDA -- PI: AM). The results presented here reflect the authors' views; the ERC is not responsible for their use. LVEK acknowledges support from a SKA-NL Roadmap grant from the Dutch ministry of OCW.

\bibliography{Papers}

\end{document}